# Component-wise dimensionally reduced flows and helicity conservation

Jian-Zhou Zhu (朱建州)[a]

*Su-Cheng Centre for Fundamental and Interdisciplinary Sciences, China*

The component-wise dimensionally reduced real Schur flows (RSFs) associated to the classical compressible Euler equation [J.-Z. Zhu, J. Math. Phys. **62**, 083101 (2021)] is reformulated alternatively in terms of mode-truncation, with the untruncated Fourier modes preserving the original interaction structure and thus other important derivatives. A number of results are set up for the mathematical physics of component-wise dimensionally reduced flows (CWDRFs, including those with further dimensional reductions of RSFs); and, it is particularly shown that previous proofs of the helicity invariance in barotropic ideal flows were overkilling in the sense of using the unnecessary condition of local mass conservation, while our new "sharper" proof without invoking the latter carries over to our CWDRFs and the inviscid Burgers equation, verified using recent results [S. G. Chefranov & A. S. Chefranov, Phys. Scr. **94**, 054001 (2019)] for the latter case in the infinite domain.

[a] Electronic mail: jz@sccfis.org

## I. INTRODUCTION

Modeling (highly) compressible systems with strong constraints, beyond Julien & Knobloch,[1] say, can benefit from studying the component-wise dimensionally reduced flows (CWDRFs) which are defined to be those whose velocity $\boldsymbol{u}$ has the gradient $\nabla\boldsymbol{u}$ with some common component(s) uniformly vanishing over space and time. The general flows, such as one-dimensional (1D), fully two-dimensional (2D) or fully three-dimensional (3D) flows can be regarded as special cases with the same dimensional reduction for all velocity components, while the so-called real Schur flows[2–4] are CWDRFs, with, for instance, only two or one component of the three subject to dimensional reduction.[2]

We will be working in the 3-dimensional Euclidean space $\mathbb{E}^3$, so the relevant basic knowledge of matrix theory in any standard text books[5] is introduced as follows:

**Definition 1** *In* $3\times 3$ *matrix* $R=\{r_{ij}\}$*, if all other elements are arbitrary real numbers except that* $r_{31}=r_{32}=0$*, then* $R$ *is called the 223-type real Schur matrix; if all other elements are arbitrary real numbers except that* $r_{13}=r_{23}=0$*, then* $R$ *is called the 331-type real Schur matrix; if both are satisfied,* $R$ *is the 221-type (real Schur matrix).*

**Proposition 1** *All* $3\times 3$ *real matrices can be turned into the 223- and 331-type real Schur ones by orthogonal (real Schur) transformations.*

Now since $\nabla\boldsymbol{u}(\boldsymbol{x},t)$ can always be locally (in space $\boldsymbol{x}$ and time $t$) transformed into the real Schur form, the corresponding CWDRFs — real Schur flows (RSFs) — based on such consideration of the "genericity" may be reminiscent of the local inertial frame and its role in connecting the special and general relativity theory and might be of fundamental importance. More specifically, CWDRFs are characterized by the uniformly vanishing of some component(s) of the velocity gradient matrix $\{u_{i,j}\}$ of the velocity $\boldsymbol{u}=\{u_1,u_2,u_3\}=\{\boldsymbol{u}_h,u_3\}$ ("$_h$" for the *horizontal plane* and "$_3$" for the *vertical direction*),

$$G=\begin{pmatrix} u_{1,1} & \boxed{u_{2,1}} & \cancel{u_{3,1}} \\ \enclose{circle}{u_{1,2}} & u_{2,2} & \cancel{u_{3,2}} \\ \cancel{u_{1,3}} & \cancel{u_{2,3}} & u_{3,3} \end{pmatrix}, \tag{1}$$

where an index behind the comma denotes the spatial derivative with respect to the corresponding coordinate variable. Two categories of CWDRFs are of particular interest due to their special basic properties:

- One is that with vertically invariant horizontal velocity $\boldsymbol{u}_h$, $\partial_{x_3}\boldsymbol{u}_h \equiv \boldsymbol{0}$, i.e.,

$$u_{1,3} \equiv 0 \equiv u_{2,3} \tag{2}$$

(2D2D3D, hence termed 223RSF) as indicated by the multiple slashes, or that with horizontally invariant vertical velocity $u_3$, $\nabla_h u_3 \equiv \boldsymbol{0}$, i.e.,

$$u_{3,1} \equiv 0 \equiv u_{3,2} \tag{3}$$

(3D3D1D, hence termed 331RSF) as indicated by single slash in the above matrix (1). Obviously, the 'intersection" of these two types of RSFs are the 2D2D1D CWDRFs.

- The other has, in Eq. (1),

$$u_{1,2} \equiv 0 \text{ (circled) besides } \partial_{x_3}\boldsymbol{u}_h \equiv \boldsymbol{0} \tag{4}$$

(1D2D3D), or has

$$u_{2,1} \equiv 0 \text{ (boxed) besides } \nabla_h u_3 \equiv \boldsymbol{0} \tag{5}$$

(3D2D1D). 1D2D3D or 3D2D1D CWDRFs are uniformly free from complex eigenvalues, with all real eigenvalues being the diagonal elements of $G$ at each location, thus may be called 'lone' Schur flow (LSF).

Obviously, we have

**Proposition 2** *The rotation matrix $Q$ around the $x_3$ axis is of 221-type and transforms the 331RSF and 223RSF into the respective same type; that is, $Q$ preserves the zero patterns of 331RSF and 223RSF.*

**Remark 1** *The above rotation generally does not preserve the zero patterns of the correspondingly more special LSF structure, which is connected with the exclusion of closed streamlines (rotational symmetry) — Theorem 3 to be proved later.*

While previous literatures had used only (some special form of the) local 223-type real Schur matrix in hydrodynamic studies, Ref. 2 exploited the genericity of the general real Schur matrices [partly with the physical motivation associated to the analogy of the theoretical framework of *general relativity* concerning *local inertial frame* and *(strong) equivalence principle*] and examined both the 331 and 223RSFs (c.f., e.g., Refs. 6 and 7 for comprehensive bibliographies of fluid literatures relevant to real Schur matrices.)

It has been noticed[3] that a formal Taylor-Proudman fast-rotating limit of a barotropic flow happens to correspond to 223RSF, except that the former is additionally incompressible for the flow velocity components in the rotation plane. So far, it is not very clear to which flow situation the 3D3D1D CWDRF, i.e., the 331RSF precisely corresponds, while the topology, characterized by the equilibrium plane of the 1D velocity component (see below), somehow resembles the flows with (multiple) infinitely thin stratosphere(s), Venus' multiple temperature inversion layers, and pycnocline(s) which hinder the vertical convection (thermoclines and haloclines in the ocean also can do so, depending on the situations); but, Ref. 2 also mentioned the possibility of active flows: Self-propelled particles in nonequilibrium fluids, such as bacterial suspensions or artificial microswimmer solutions, autonomously consume energy to generate motion, which results in distinctive collective phenomena, including spontaneous flow, vortex formation, and even turbulent dynamics, and which might realize whatever CWDRFs we want. Probably a good understanding of it, with the help of numerical simulations, will in turn help identify the relevant physical systems. While Ref. 3 studies also higher-dimensional flows and nonbarotropic 223RSF structures, here we will focus on barotropic CWDRFs, but on the other hand, will explore deeper and wider.

For CWDRFs, associated to the Euler equation, say, with a particular type of dimensional reduction, there is probably no completely universal method or (reduced) model for them. The Euler equation is universal in the sense that various flows are realized with different forces and initial and boundary conditions but with the same equation, however there may be different ways leading to and maintaining that dimensional reduction, resulting in nonuniversal models. Nevertheless, it still makes a lot of sense to find a self-consistent model for CWDRFs that can be used as a fundamental template (it turns out that our model is a Fourier-mode truncation of the Euler equation and has the same mode-interaction structure as the original system for the untruncated modes, thus of particular fundamental role) for theoretical studies and for simulating and understanding realistic situations, as will be clarified in the following sections: Sec. II presents the various CWDRF models mentioned in the above and proves the inequivalence and equivalence, respectively, between the two types in the first RSF and the second LSF categories; and, Sec. III proves theorems about the existence of closed streamlines; while, the invariant sub(manifold), and invariace of helicity are proven in Sec. IV; Sec. V is for the concluding remarks. The appendices bear some of the proofs and computational details from the main text.

## II. MODELS AND UNIQUENESS

Consider the Navier-Stokes equation with density $\rho$, pressure $p$, damping $\boldsymbol{d}$ and acceleration $\boldsymbol{a}$,

$$\partial_t \rho + \nabla \cdot (\rho \boldsymbol{v}) = 0, \tag{6}$$

$$\partial_t \boldsymbol{v} + \boldsymbol{v} \cdot \nabla \boldsymbol{v} = -\rho^{-1} \nabla p + \boldsymbol{d} + \boldsymbol{a}. \tag{7}$$

For simplicity, the damping here is chosen to be $\boldsymbol{d}(\boldsymbol{v}) = \nu \nabla^2 \boldsymbol{v}$ with constant kinetic viscosity $\nu$, generally applied in incompressible flows. When the pressure term is absent, we have the Burgers equation. We actually will be focusing on the CWDRFs associated to the ideal barotropic Euler equation (free of dissipation $\boldsymbol{d}$ and forcing $\boldsymbol{a}$), while the extensions are obvious.

As before for the RSFs,[3,4] CWDRFs are supposed to live in the cyclic box of dimension $2\pi$ (the problem in the infinite domain with the fields decaying fast enough at the infinity can be similarly analyzed but is beyond the purpose here). Spatial averages with $\langle \bullet \rangle_J$, where the index (tuple) $J$ refers to the coordinate axis(es) to be integrated over, will be used to facilitate the dimensional reduction; e.g., among others,

$$\langle \bullet \rangle_{123} := (2\pi)^{-3} \int \int \int \bullet d^3 \boldsymbol{x}, \, \langle \bullet \rangle_{12} := (2\pi)^{-2} \int \int \bullet dx_1 dx_2. \tag{8}$$

And for further simplification, we will let the pressure $p$ follow the adiabatic relation

$$p = c^2 \rho, \, \rho = \rho_0 e^{\zeta}, \tag{9}$$

with $c$ being the sound speed and $\rho_0$ a background density which can be taken to be unit (with appropriate normalization), thus $\zeta = \ln \rho$, for convenience. So, we have the Euler equation

$$\partial_t \zeta + \zeta_{,\sigma} u_\sigma + u_{\alpha,\alpha} \; = \; 0, \tag{10}$$

$$\partial_t u_\lambda + u_\sigma u_{\lambda,\sigma} + c^2 \zeta_{,\lambda} \; = \; 0, \tag{11}$$

with $(\bullet)_{,\gamma} = \partial(\bullet)/\partial x_\gamma$ and the Einstein summation convention over repeated indexes, starting from which a self-consistent set of RSF equations can be constructed.[3,4]

The derivation in Ref. 3 for 223RSF can be carried over to 331RSF and to LSF, but, to be self-contained and to offer a different anlge of view, we will also offer an alternative formulation and explanation in terms of Fourier-mode *truncation* (or *selection* with the *inclusion-exclusion principle*) preserving the interaction structure of the original compressible Euler equation.

### A. RSF

**Lemma 1** *The RSF velocity gradient matrix is preserved as the same real Schur form by the self-advection term $\boldsymbol{u} \cdot \nabla \boldsymbol{u}$.*

*Proof.* It is direct to verify that $\partial_3(\boldsymbol{u} \cdot \nabla \boldsymbol{u}_h) \equiv 0$, with Eq. (2), thus no source of $\partial_3 \boldsymbol{u}_h$ for the deviation from the 223RSF and $\nabla_h(\boldsymbol{u} \cdot \nabla u_3) \equiv \boldsymbol{0}$, with Eq. (3), thus no source of $\nabla_h u_3$ for the deviation from the 331RSF. □

**Remark 2** *This means the ideal Burgers equation (inviscid and without the acceleration term) admits RSFs constituting the invariant manifold, but we will be considering the more general compressible Euler equation.*

With barotropic pressure, the above Lemma 1 means the momentum equation in the Euler equation can be written as the following explicitly reduced forms:

$$\partial_t \boldsymbol{u}_h + \boldsymbol{u}_h \cdot \nabla \boldsymbol{u}_h = -c^2 \nabla_h \zeta, \tag{12a}$$

$$\partial_t u_3 + \boldsymbol{u} \cdot \nabla u_3 = -c^2 \zeta_{,3} \tag{12b}$$

for the 223RSF and

$$\partial_t \boldsymbol{u}_h + \boldsymbol{u} \cdot \nabla \boldsymbol{u}_h = -c^2 \nabla_h \zeta, \tag{13a}$$

$$\partial_t u_3 + u_3 u_{3,3} = -c^2 \zeta_{,3} \tag{13b}$$

for the 331RSF, given accordingly the two types of initial fields. To preserve the respective global RSF structures, we should have

$$\nabla_h(\zeta_{,3}) \equiv 0 \text{ or } k_h k_3 \hat{\zeta}_{\boldsymbol{k}} \equiv 0. \tag{14}$$

The latter in Fourier space means that for the wavevector $\boldsymbol{k}$ with nonvanishing multiplication of its horizontal component $k_h$ and vertical component $k_3$, $k_h k_3 \neq 0$, such $\zeta$-modes should be truncated or reduced, *i.e.*, accordingly the Fourier coefficient $\hat{\zeta}_{\boldsymbol{k}} \equiv 0$ being imposed; that is, the untruncated modes are those of $k_h = 0$ or $k_3 = 0$, corresponding to the wavevector subsets $S_J$ with $J$ being respectively the index (tuple) $h$ and 3. Obviously, with '$h$'↔'12' for the index tuple and $S_{123}$ for set $\{\boldsymbol{0}\}$, i.e., for $k_1 = k_2 = k_3 = 0$,

$$S_{12} \cap S_3 = S_{123}. \tag{15}$$

So, we have the untruncated-mode wavevector set

$$S = S_{12} \cup S_3, \text{ with } |S| = |S_{12}| + |S_3| - |S_{12} \cap S_3| = |S_{12}| + |S_3| - |S_{123}| \quad (16)$$

by the so-called *inclusion-exclusion principle*. Since the vanishing of the wavevector component corresponds accordingly to the spatial average, we have

$$\zeta = \langle\zeta\rangle_3 + \langle\zeta\rangle_{12} - \langle\zeta\rangle_{123}, \quad (17)$$

with the spatial averages $\langle\bullet\rangle_J$ having the same index (tuple) $J$ in $S_J$. Then, with the interchangeability of the time derivative and spatial average and $\varrho = \boldsymbol{u} \cdot \nabla \ln \rho + \nabla \cdot \boldsymbol{u}$ from the continuity Eq. (10), the latter becomes

$$\partial_t \zeta = \langle\varrho\rangle_{123} - \langle\varrho\rangle_3 - \langle\varrho\rangle_{12}. \quad (18)$$

So, Eqs. (18,12a,12b) and Eqs. (18,13a,13b) govern respectively the 223RSF and 331RSF associated to the barotropic Euler equation. The construction means that *they share the same interaction structure as the original Euler equation for the untruncated modes*, indicating the inheritance from the Euler equation of relevant derivatives (such as the Liouville theorem) for other analyses, such as some relevant analytical theories of turbulence (when other ingredients such as dissipation are appropriately included) and closure models, which is not the purpose here but on which we will come back to briefly remark in Sec. V.

The same Fourier-space interaction structure of 223RSF and 331RSF $\zeta$-modes, alternative to the same variable seperation [Eq. (31) below] used in Ref. 3 to derive Eq. (17) in the configuration space, does not assure that the two types of RSFs are the same, since the mode truncations in their respective momentum equations, Eqs. (13a, 13b) versus Eqs. (12a,12b), are obviously different.

So, given the mathematical fact that every real matrix can be orthogonally transformed into either of the two real Schur forms, it is eligible to ask: are the 223RSF and 331RSF really different, or they are actually “the same” after an orthogonal coordinate transformation (real Schur transformation)? We notice that either of the RSFs have the velocity gradients uniformly over space and time of a particular real Schur form but with (dynamical) variations of the non-zero elements over space and time, so if they are the same, we should have a uniform transformation to turn the varying real Schur form velocity gradient matrix, say, for the 223RSF, into the other real Schur form, say, for the 331RSF. However, this is not possible by the following “no-go” theorem:

**Theorem 1** *There is no uniform orthogonal (real Schur) transformation to turn all 331-type real Schur matrices into the 223-type.*

*Proof.* According to Definition 1, let $T = \{\tau_{i,j}\}$ be the 331-type real Schur matrix, with $\tau_{13} = \tau_{23} = 0$, and $S = \{s_{i,j}\}$ be 223-type, with $s_{31} = s_{32} = 0$, leaving 7 elements being arbitrary real numbers for either type. Then suppose $Q = \{q_{ij}\}$ be a $3 \times 3$ real matrix, satisfying the orthogonality $Q^\top Q = I$. Straightforward but tedious calculation (see Appendix A for the details) from $S = Q^\top T Q$, with 331-type $T$ of seven arbitrary real elements, for the elements $s_{31} = 0$ and $s_{32} = 0$ (leading to $2 \times 7 = 14$ quadratic monomial equations of $q_{ij}$) shows that the conditions which $Q$ should satisfy are in contradiction with the orthogonality assumption. □

**Remark 3** *The statement of this theorem already implies its "mirror image"; that is, "there is no uniform orthogonal (real Schur) transformation to turn all 223-type real Schur matrices into the 331-type", whose transformation, if existed, would be the inverse of the $Q$ in the above proof.*

So, 223RSF and 331RSF are indeed "different", which however is not the case for LSF by Proposition 3 in the next Sec. II B: we have only one type of LSFs, and we need and will only study the 1D2D3D CWDRFs which can be called LSF without ambiguity.

## B. LSF

For the LSF satisfying Eq. (4) and associated to the aforementioned Euler equation in a periodic box (or similarly in the infinite domain with fields decaying sufficiently fast at the infinity), we should have

$$\zeta_{,12} = \zeta_{,13} = \zeta_{,23} \equiv 0 \tag{19}$$

to preserve the LSF structure in the momentum equation

$$\partial_t u_1 + u_1 u_{1,1} = -c^2 \zeta_{,1}, \tag{20a}$$

$$\partial_t u_2 + \boldsymbol{u}_h \cdot \nabla_h u_2 = -c^2 \zeta_{,2}, \tag{20b}$$

$$\partial_t u_3 + \boldsymbol{u} \cdot \nabla u_3 = -c^2 \zeta_{,3}; \tag{20c}$$

Eq. (19 written in the Fourier space goes as follows

$$k_1 k_2 \hat{\zeta}_{\boldsymbol{k}} = k_1 k_3 \hat{\zeta}_{\boldsymbol{k}} = k_2 k_3 \hat{\zeta}_{\boldsymbol{k}} \equiv 0, \text{ i.e., } \hat{\zeta}_{\boldsymbol{k}} \equiv 0 \text{ unless } k_1 k_2 = k_1 k_3 = k_2 k_3 = 0. \tag{21}$$

As in the 223- and 331-RSF case, such a configuration pattern corresponds to the inclusion-exclusion principle for modes in wavevector sets, which means that, again, for the untruncated modes, such CWDRFs actually have the same interaction structure as the original Euler equation, and thus the relevant derivatives: Note that the wavevector sets (again $S_J$, such as that with the tuple $J = 12$ for untruncated-mode subset of $k_1 = k_2 = 0$, corresponding to the spatial average $\langle\bullet\rangle_J$ below with the same index tuple $J$) satisfy

$$S_{12} \cap S_{13} = S_{12} \cap S_{23} = S_{13} \cap S_{23} = S_{123}. \tag{22}$$

So, now for the untruncated-mode wavevector set $S$, corresponding to Eq. (21) in the above or (26) below,

$$\begin{aligned} S &= S_{12} \cup S_{13} \cup S_{23}, && (23) \\ |S| &= |S_{12}| + |S_{13}| + |S_{23}| - |S_{12} \cap S_{13}| - |S_{12} \cap S_{23}| - |S_{13} \cap S_{23}| + |S_{123}| && \\ &= |S_{12}| + |S_{13}| + |S_{23}| - 2|S_{123}|. && (24) \end{aligned}$$

With the vanishing of the wavevector component corresponding accordingly to the spatial average,

$$\zeta = \langle\zeta\rangle_{23} + \langle\zeta\rangle_{13} + \langle\zeta\rangle_{12} - 2\langle\zeta\rangle_{123}, \tag{25}$$

and with the interchangeability with time derivative of the latter, we have

$$\partial_t \zeta = 2\langle\varrho\rangle_{123} - \langle\varrho\rangle_{23} - \langle\varrho\rangle_{13} - \langle\varrho\rangle_{12}. \tag{26}$$

Alternatively, similar consideration of function factorization as in Ref. 3 for the RSF, the above Eq. (26) is from the requirement that, to preserve the LSF structure, Eq. (19) be satisfied, thus the decomposition of $\zeta$ into three spatially univariate functions, $\zeta = \mathscr{P}_3(x_3, t) + \mathscr{P}_2(x_2, t) + \mathscr{P}_1(x_1, t)$, thus Eqs. (25 and 26).

Eqs. (26,20a,20b,20c) constitute the LSF system associated to the barotropic Euler equation.

As noted earlier, the LSF is unique in the sense that we have

**Proposition 3** *All upper triangular* $3\times3$ *real matrices can be transformed into the lower triangular real matrices by a uniform orthogonal transformation.*

*Proof.* In the same calculation of the proof for Theorem 1 but now with one more vanishing element, such orthogonal $Q$ exists: actually we have a simplest example of anti-diagonal permutation matrix which switches the $x_1$ and $x_3$ axes and thus the corresponding triangular matrixes (for whatever values of the other nonvanishing elements). □

## III. TOPOLOGY

Theory of ordinary differential equations[8] often employ the following lemma implicitly:

**Lemma 2 (monotonicity of one-dimensional autonomous system)** *For the one-dimensional autonomous system*

$$\frac{dy}{ds} = f(y),$$

*where $f : \mathbb{R} \to \mathbb{R}$ is a continuous function, if $f(y) \neq 0$ in some interval $I \subseteq \mathbb{R}$, then the solution $y(s)$ on $I$ is strictly monotonic.*

The proof of the above lemma is almost trivially equivalent to stating the property of continuity, thus omitted here and rarely stated in other texts explicitly, but appropriate formulation of the problems and nuanced adjustments are needed when going beyond one-dimensional systems, such as discussing how the 331 and 223RSFs are different topologically:

**Theorem 2** *331RSFs can have closed streamlines only in the horizontal equilibrium plane of $u_3$, i.e., the $x_1$-$x_2$ plane at $x_3 = x_3^{\#}$ where $u_3(x_3^{\#}) = 0$.*

*Proof.* 331RSFs have $u_3 = u_3(x_3, t)$. At any given time $t$, $dx_3/ds = u_3(x_3)$ on the streamline parametrized by $s$ is a one dimensional autonomous system, then, according to Lemma 2, $x_3$ is monotonic at nonequilibrium points. If we have the equilibrium point at $x_3^{\#}$ approaching which from one side $x_3$ is monotonically increasing, and if it is monotonically decreasing on the other side, then the system terminates at $x_3^{\#}$; *vice versa*. This means that $x_3$ on the streamline can only go monotonically or stay at $x_3 = x_3^{\#}$ where $u_3(x_3^{\#}) = 0$, with no possibility of being periodic. So, any closed streamlines can only be on the $x_1$-$x_2$ equilibrium-plane. □

**Remark 4** *The 223RSF allows non-monotonicity of $u_3$, which appears particularly clear in the helical cases as can be seen in some figures presented in Ref. 4. It might be that 331RSF is closer to the stratification physics, while 223RSF closer to rotation. Indeed, while the Earth in general has only one main stratosphere, in Venus' atmosphere, the absorption of solar radiation by sulfuric acid clouds creates multiple temperature inversion layers, which, like the $x_1$-$x_2$ equilibrium plane where $x_3 = x_3^{\#}$ and $u_3(x_3^{\#}) = 0$, repeatedly suppress vertical convection, resulting in a much more complex vertical structure compared to Earth's: Venus rotates extremely slowly, with its day being longer than its year.*

The above topological results can be easily tested by writing down the arbitrary functions of the corresponding $\boldsymbol{u}$ and by checking the dynamical RSFs at any time, which should also be helpful in making physical connections by matching the patterns before comparing the dynamical mechanism for final examination. Actually, Theorem 2 (together with its physical connections accordingly speculated in Remark 4) and Theorem 3 below were conjectured from numerical observations.[11] For instance, Figure 1 displays typical streamlines from solutions of the 331RSF equations in a periodic domain (adapting the same numerical method and procedure in Ref. 4), where the perspective and projection views, with $x$, $y$ and $z$ respectively for our $x_1$, $x_2$ and $x_3$: the number of equilibrium planes (corresponding to physical "layers", if indeed) of $u_3$ depends on the initial data and the time of the snapshot, and its steady state can be completely determined by the driving (if it exists).

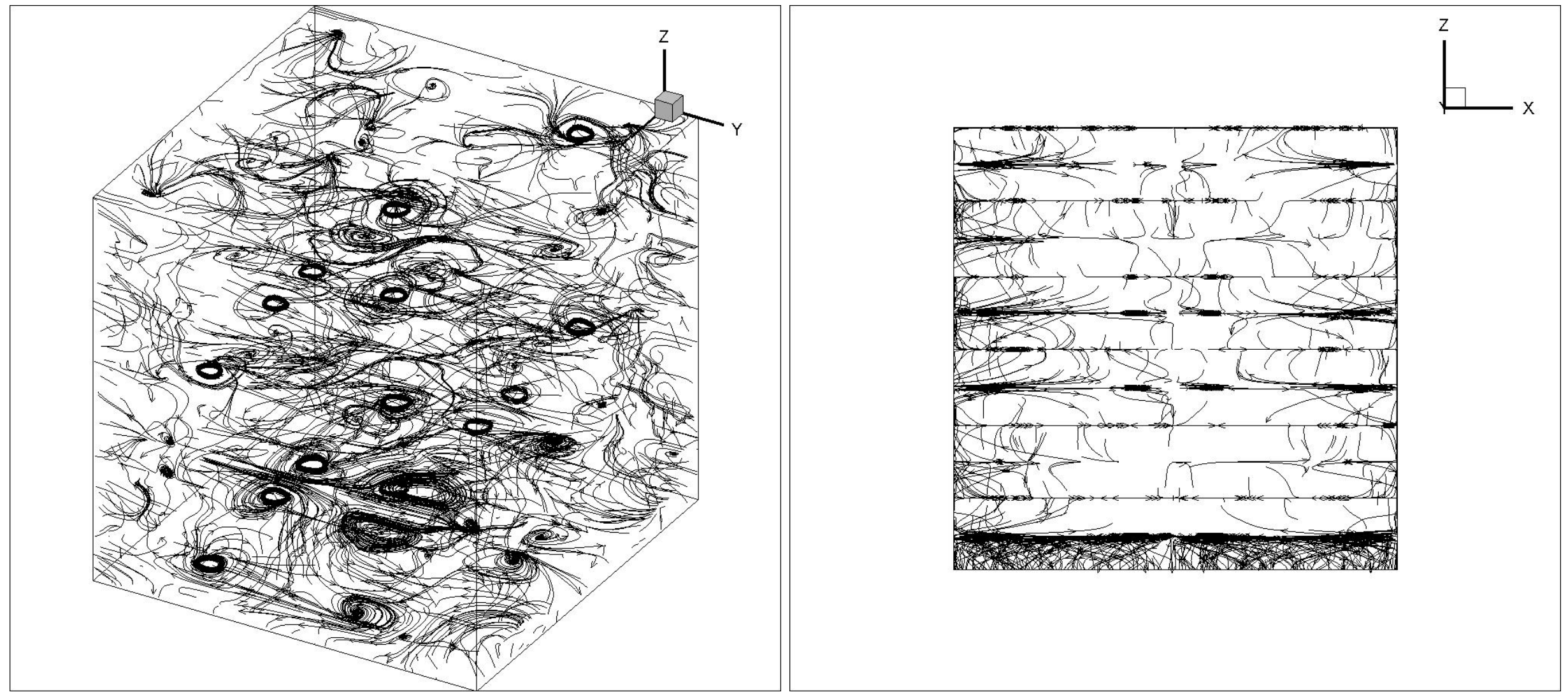

FIG. 1. 331RSF streamlines: perspective (left) and projection (right) views.

**Theorem 3** *LSFs cannot have closed streamlines.*

*Proof.* The 1D2D3D characteristics of LSFs mean that $u_1 = u_1(x_1, t)$, $u_2 = u_2(x_1, x_2, t)$ and $u_3 = u_3(x_1, x_2, x_3, t)$. Consider the instantaneous streamline parametrized by $s$:

$$\frac{dx_1}{ds} = u_1(x_1) \tag{27}$$

$$\frac{dx_2}{ds} = u_2(x_1, x_2) \tag{28}$$

$$\frac{dx_3}{ds} = u_3(x_1, x_2, x_3) \tag{29}$$

Suppose we have a closed streamline $\mathscr{C}$ with coordinate $\mathbf{x}(s)$, then we can apply the previous argument, in Theorem 2 for the $x_3$ coordinate there, subsequently to our $x_1$, $x_2$ and $x_3$ here: The coordinates on $\mathscr{C}$ should be periodic, so, if $x_1$ in Eq. (27) is not on the equilibrium point, then contradiction with the monotonicity of Lemma 2 follows; that is, $x_1$ can only take its equilibrium value $C_1$, i.e., $x_1 \equiv C$. Once $x_1$ is fixed, Eq. (28) for $x_2$ effectively becomes a univariate function of $x_2$, the latter thus being a constant for the same reason; further more, $x_3$ should also take the constant value with $x_1$ and $x_2$ fixed. With all its coordinates being constants, $\mathscr{C}$ cannot be a nontrivial closed streamline. □

**Remark 5** *The argument of monotonicity of one-dimensional autonomous system superficially fails for a periodic domain, but a streamline going repeatedly from one side of the cyclic box to the other side does not essentially constitute a closed streamline in the common sense.*

Like many terminologies in other disciplines, here it appears that, due to the historical reason and the daily language usage, "ir/rotational flow", "vortex", "vorticity" and "swirl" have been used with great ambiguity, with various proposals of defining or identifying spatially extended vortex (Refs. 6 and 7 and references therein), but, based on the above theorems concerning closed streamlines, we now may introduce some "normalization of the terminologies" with the following

**Definition 2** *"Vortex (structure)" is some distribution of vorticity $\omega = \nabla \times \boldsymbol{u}$. As for the choice of the distributions, it should be dependent on the purpose. Any other "structures" not defined by vorticity should not be called "vortex (structure)". And, a closed streamline defines the "swirl".*

**Remark 6** *The vorticity of LSFs is in general non-zero (non-potential, except for further reduction), thus "vortical but with no swirls".*

We remark that the theorems proved in this section has nothing to do with the dynamics, so they can be applied to quantum flows and higher (than 3) dimensional theories with similar structures.

## IV. DYNAMICS

### A. Invariance of helicity

The helicity of a flow field $\boldsymbol{u}$,

$$\mathcal{H} = \langle \nabla \times \boldsymbol{u} \cdot \boldsymbol{u} \rangle_{123}, \tag{30}$$

has been known to be dynamically conserved by the ideal flows governed by the incompressible and compressible barotropic Euler equation. The compressible barotropic case is relevant to our CWDRFs here, but, to the best of our knowledge, the proofs (popularly) accepted in the community need the mass conservation (continuity) equation: c.f., Moreau[9] and Khesin & Chekanov[10]. Such a status is too restrictive and may be misleading, in the sense that the traditional local form of mass conservation is not a necessary condition for the helicity conservation which thus can be extended to broader systems as explained in Appendix B.

We will establish theorems which claim that Burgers and barotropic ideal flows conserve the helicity without the necessity of invoking mass conservation, and as a corollary, helicity is also conserved by the RSFs associated to either of them.

**Theorem 4** *A 3D Burgers (pressureless) ideal flow conserves the helicity.*

There are different ways of calculations, including the traditional vector calculus and that using the differential forms, in Lagrangian or Eulerian descriptions, to prove this theorem. It can be completed by working directly with the inviscid Burgers equation or by simple reduction from the case of barotropic compressible ideal flows in the following

**Theorem 5** *Helicity is conserved by an ideal barotropic flow, which can be proved without invoking the mass conservation.*

The proof is given in Sec. B 1 of Appendix B. And, as a further specialization of the "extended" helicity conservation law as remarked in Sec. B 2, we then have

**Corollary 1** *The ideal RSF and LSF, associated to either the Burgers or Euler equation, conserve helicity.*

**Remark 7** *The conservation of helicity by the Burgers ideal dynamics can also be directly verified by the exact Lagrangian mapping form of the solutions in the infinite domain given by Chefranov & Chefranov[12] (see also Chefranov[13]), as presented in the Appendix B 2.*

## B. Invariant submanifold

The fact that pressureless Burgers, i.e., the self-advection in the Euler equation, preserves the RSF structure means the RSF invariant manifold, so in some sense proper analogy with the flat space-time of special relativity might be helpful, say, for problems such as what kind of tools are needed to reformulate classical hydrodynamics.

We have also noticed that under extreme conditions or strong constraints, component-wise dimensional reduction can happen. But, we can also ask: "Does the 'free (undriven)' classical hydrodynamic equation have the relevant invariant manifold?" Ref. 4 presents results that show the differences of the flows, starting from the same initial data, between the classical fluid equation and the RSF equation, so, from such a distinct situation compared to the Burgers case (in the above analogy with the globally flat space-time), it seems that we might associate, by carrying even further the analogy, the pressure term with the mass in the Einstein equation, the latter causing the curving of space-time.

With the hope of progress beyond superficial analogy, here we dig a bit deeper into the details. In constructing the "free" RSF equation,[3,4] we have required the pressure term to also preserve the RSF structure, i.e., the decomposition, corresponding to Eq. (16) for the untruncated modes of wavevector set $S$,

$$\zeta_{,31} \equiv 0 \equiv \zeta_{,32} : \ \zeta = \mathscr{Z}_3(x_3, t) + \mathscr{Z}_h(\boldsymbol{x}_h, t). \tag{31}$$

So, if the RSF solution also satisfies the original Euler equation, we should have the "compatibility" condition:

$$(\nabla_h \varrho)_{,3} \equiv 0. \tag{32}$$

That is, we should also have the decomposition in the continuity equation (10):

$$\varrho = \mathscr{R}_h(\boldsymbol{x}_h, t) + \mathscr{R}_3(x_3, t) \iff \varrho = \langle\varrho\rangle_{12} + \langle\varrho\rangle_3 - \langle\varrho\rangle_{123}. \tag{33}$$

A special case which satisfies the compatibility condition is that satisfying both Eqs. (2) and (3). These are the 2D2D1D CWDRFs constituting the intersection of the 331RSF and 223RSF sets [the compressible, barotropic analog of the strongly rotating stratified flows studied in Ref. 1 might be made loosely, though the comparison is only formal], which indicates:

$$\mathscr{R}_h(\boldsymbol{x}_h) = \boldsymbol{u}_h \cdot \nabla_h \mathscr{Z}_h / c^2 + \nabla_h \cdot \boldsymbol{u}_h, \tag{34}$$

$$\mathscr{R}_3(x_3) = u_3 \mathscr{Z}_{3,3} / c^2 + u_{3,3}. \tag{35}$$

Correspondingly, the Euler equation admits such 2D2D1D CWDRF invariant manifold. For the initial $u_1 = \cos x \sin y$, $u_2 = -\sin x \cos y$, $u_3 = \sin x \sin y \cos z$, and constant $\rho$, we have $\varrho = -\sin x \sin y \sin z$, allowing no decomposition into $\mathscr{R}_h$ and $\mathscr{R}_3$, which leads to the different evolutions shown in Fig. 5 of Ref. 4. So, we reiterate the following

**Theorem 6** *If the RSF solves the Euler equations (10,11), it should satisfy the compatibility condition (32); and, particularly the 2D2D1D CWDRFs constitute the invariant manifold of the Euler equation.*

*Proof.* The remarks before the statement of the theorem already explain that the formula (31), (2) and (3) lead to Eq. (33). Substituting the latter into (10), and averaging over $x_3$ or $\boldsymbol{x}_h$, we see that Eq. (31) is still preserved, so that the 2D2D1D structure of the CWDRFs also preserved. □

Numerical experiments within and beyond Ref. 4 show that, many other fields like Eqs. (18) and (20) there, i.e., $\boldsymbol{u} = \{\cos x \sin y, -\sin x \cos y, \sin x \sin y\}$ and $\rho(t_0) = \rho_0(1 + \epsilon \cos z)$, satisfy the compatibility condition (32) but are not on the RSF invariant manifold. In other words, not all RSF orbits are in the solution space of the classical hydrodynamic equation, viz, we have the obvious

**Proposition 4** *Not all RSFs constitute the invariant manifold of the Euler equation.*

## V. CONCLUDING REMARKS

It appears that the CWDRFs as the "ideal models" of anisotropic flows offer the convenient templates for theoretical studies of the relevant physics, including flow topology and statistical mechanics, among others.

It is the examination of the possibility of helicity invariance in ideal CWDRFs that led to the discovery of "overkill" in the previous proofs for ideal compressible barotropic flows. The CWDRF helicity invariance and the same mode-interaction structure, thus the same Fourier-space dynamics on the wavevector support as the original Euler equation (Secs. II A and II B), especially the Liouville theorem and the aeroacoustic-analogy form, among others, will make it possible to carry out the statistical mechanical analysis which can be used to understand the anisotropic turbulence noise (to be communicated elsewhere).

Also, by studying CWDRFs associated to the Euler equation, we actually discovered the 2D2D1D invariant manifold of the latter. Unlike the Burgers case, not all RSFs constitute the invariant manifold of the Euler equation; or, in other words, the pressure term coupling the rest of the fluid system leads to the fact that the real Schur transformation of the velocity gradient matrix can only be local: if we analogously borrow the language used in Einstein's gravitation theory for reference, this transformation removes the "pressure/gravity effect" in the local "real Schur/inertial" frame, which, going beyond the superficial analogy, actually raises the question of the possibilities of reducing general flow physics to that of RSFs and even of reformulating fluid mechanics. [Arnold and Khesin[14] wrote that the gauge groups in physics "occupy an intermediate position between the rotation group of a rigid body and the diffeomorphism groups [of hydrodynamics]... but are yet too simple to serve as a model for hydrodynamics." Now, the fact that we have for the local structures the "gauge similarity" ('similar' up to the local real Schur transformations of the velocity gradient) encourages attacking the problem of fundamental (statistical) hydrodynamics with the possibility of some reduction of the diffeomorphism groups as such.] The new formulation, if it exists, would be "downward", based on the genericity of the intrinsic local real Schur structures with dimensional reduction, and should be different from the idea of holographic fluid which goes "upward" to the Einstein equation in higher dimensional space-time and which is really related to the general relativity theory.

It is interesting to note that, in special relativity, the Poincaré group of transformations preserving the Minkovski metric actually admits both proper (positive-determinant) and improper (negative-determinant) Lorentz transformation, and the Nature laws (including parity invariance) pick the proper one.[15] Now we also have two types of RSFs, but with no reason to particularly exclude either fundamental template. The comparison or analogy may not make much sense but could be of some value for motivating the guess of the potentially right mathematical structure of the possible new formulation noted in the last paragraph, to shed light on the open mathematical problems of Euler and Navier-Stokes equations, say.

As mentioned, the fact that "swirls" live only in the horizontal equilibrium plane of $u_3$ of 331RSFs mimics the hindrance of vertical advections by the temperature inversion layers in Venus' atmosphere, the Earth's stratosphere and ocean pycno-, thermo- and halo-clines. Dynamical comparisons associated to stratifications will further clarify the usefulness of the 331RSF model with more systematic observations.

The simple Definition 2 of "swirls" by closed streamlines with conceptual clarity is made possible with Theorem 3 for the absence of closed streamlines in LSFs, and Remark 6 indicates that LSF is somehow inbetween the conventional notions of "potential" and "rotational" flows; but, so far, we have not been able to find its concrete physical connections with realities, which probably is "practically" rooted in the necessity of additional dimensional reductions beyond the generic RSF structures, requiring even more special or stronger constraints (c.f. Remark 1 and Ref.

1). The beauty in such results however is believed to be of mathematical physics value.

## ACKNOWLEDGMENTS

At this end of the beginning, the author acknowledges H. Xiang, S. Xiong and P. Xu for the interests and interactions which helped motivating some of the problems and results on CWDRFs. He also thanks S. G. Chefranov for kindly imparting Refs. 12 and 13.

## DATA AVAILABILITY

The data that support the findings of this study are available from the corresponding author upon request.

## Appendix A: The detailed proof of Theorem 1

*Proof.* For clarity, the detailed computational proof here adopts slightly different notations and presentation styles compared to the main text.

Define two types of Schur matrices:

$$T = \begin{pmatrix} * & * & 0 \\ * & * & 0 \\ * & * & * \end{pmatrix}, \quad S = \begin{pmatrix} * & * & * \\ * & * & * \\ 0 & 0 & * \end{pmatrix} \tag{A1}$$

where $*$ denotes arbitrary real numbers. For $T$, the zero elements are fixed at positions $(1,3)$ and $(2,3)$; for $S$, zeros are fixed at $(3,1)$ and $(3,2)$.

Let $Q$ be an arbitrary $3 \times 3$ orthogonal matrix with column vectors $\mathbf{q}_1$, $\mathbf{q}_2$, $\mathbf{q}_3$. For matrix $T$ above, compute the elements of $S = Q^\top T Q$:

$$S(3,1) = \mathbf{q}_3^\top T \mathbf{q}_1 = 0 \tag{A2}$$

$$S(3,2) = \mathbf{q}_3^\top T \mathbf{q}_2 = 0 \tag{A3}$$

Expanding $T\mathbf{q}_1$:

$$T\mathbf{q}_1 = \begin{pmatrix} aq_{11} + bq_{12} \\ cq_{11} + dq_{12} \\ eq_{11} + fq_{12} + gq_{13} \end{pmatrix} \tag{A4}$$

with $T(i, j)$ being replaced by $a$, $b$, $c$ etc., for clarity in a self-evident way.

From Eq. (A2), we derive the linear system:

$$q_{31}q_{11} = 0,\ q_{31}q_{12} = 0,\ q_{32}q_{11} = 0,\ q_{32}q_{12} = 0,\ q_{33}q_{11} = 0,\ q_{33}q_{12} = 0,\ q_{33}q_{13} = 0. \tag{A5}$$

Similarly, from Eq. (A3):

$$q_{31}q_{21} = 0,\ q_{31}q_{22} = 0,\ q_{32}q_{21} = 0,\ q_{32}q_{22} = 0,\ q_{33}q_{21} = 0,\ q_{33}q_{22} = 0,\ q_{33}q_{23} = 0. \tag{A6}$$

Using orthonormality $\mathbf{q}_i^\top \mathbf{q}_j = \delta_{ij}$ and systems (A5)-(A6):

1. If $q_{33} \neq 0$, then $\mathbf{q}_1$ and $\mathbf{q}_2$ must satisfy $q_{13} = q_{23} = 0$ and $q_{11} = q_{12} = 0$, leading to $\mathbf{q}_1 = \mathbf{0}$, which is a contradiction.

2. Thus $q_{33} = 0$, and the system reduces to:

$$\begin{cases} q_{31}q_{11} = q_{31}q_{12} = 0 \\ q_{32}q_{11} = q_{32}q_{12} = 0 \\ q_{31}q_{21} = q_{31}q_{22} = 0 \\ q_{32}q_{21} = q_{32}q_{22} = 0 \end{cases} \tag{A7}$$

3. Since $q_{31}^2 + q_{32}^2 = 1$, assume without loss of generality $q_{31} \neq 0$, then:

$$q_{11} = q_{12} = q_{21} = q_{22} = 0 \tag{A8}$$

   This forces $\mathbf{q}_1 = (0, 0, q_{13})^\top$ and $\mathbf{q}_2 = (0, 0, q_{23})^\top$ to be non-orthogonal, another contradiction.

This completes the proof. □

## Appendix B: Helicity conservation independent of the continuity equation

This appendix provides a proof of helicity conservation for ideal barotropic flows using exterior calculus, which does not require the mass conservation equation. This contrasts with the proofs by Moreau[9] and later by Khesin & Chekanov,[10] which explicitly used the continuity equation to simplify the volume element. The content of this appendix is independent and self-contained, with conventional notations. The cases of smooth solutions of the inviscid Burgers and RSF equations are just simple reductions, precisely because of the independence from the continuity equation, so the new proof not only appears "sharper" but also extends the known conservation law.

### 1. New "sharper" proof of the helicity conservation

Being "sharper", our new proof somehow "unifies" both Theorems 4 and 5, and, Corollary 1:

*Proof.*

*a. Exterior Calculus Formulation* Consider a three-dimensional manifold equipped with Euclidean metric. Define the velocity 1-form $u = u_i dx^i$ (lowering indices via the metric). The vorticity 2-form is $\omega = du$, and the helicity density 3-form is $H = u \wedge \omega$.

The ideal barotropic flow (with conservative body forces allowed but not shown) equation reads $\frac{D\mathbf{u}}{Dt} = -\nabla h$, $h = \int \frac{dp}{\rho}$, which in exterior calculus becomes $\partial_t u + d\left(\frac{1}{2}|u|^2\right) + \iota_u \omega = -dh$, where $|u|^2 = u_i u^i$ and $\iota_u$ denotes interior product. Thus,

$$\partial_t u = -d\left(\frac{1}{2}|u|^2 + h\right) - \iota_u \omega. \tag{B1}$$

*b. Time Derivative of Helicity 3-form* We now compute

$$\partial_t H = (\partial_t u) \wedge \omega + u \wedge (\partial_t \omega). \tag{B2}$$

Taking exterior derivative of the expression for $\partial_t u$, we obtain the vorticity evolution:

$$\partial_t \omega = d(\partial_t u) = -d(\iota_u \omega). \tag{B3}$$

Substituting Eq. (B1) and (B3) into Eq. (B2), we have

$$\begin{aligned} \partial_t H &= \left[-d\left(\frac{1}{2}|u|^2 + h\right) - \iota_u \omega\right] \wedge \omega - u \wedge d(\iota_u \omega) \\ &= -d\left(\frac{1}{2}|u|^2 + h\right) \wedge \omega - (\iota_u \omega) \wedge \omega - u \wedge d(\iota_u \omega). \end{aligned} \tag{B4}$$

Since $d\omega = 0$, the first term is an exact form: $-d\left(\frac{1}{2}|u|^2 + h\right) \wedge \omega = -d\left[\left(\frac{1}{2}|u|^2 + h\right)\omega\right]$. For the remaining terms, use the identity $u \wedge d(\iota_u \omega) = d(u \wedge \iota_u \omega) - du \wedge \iota_u \omega = d(u \wedge \iota_u \omega) - \omega \wedge \iota_u \omega$. Noting that for a 1-form $\iota_u \omega$ and a 2-form $\omega$, the wedge product commutes: $\omega \wedge \iota_u \omega = (\iota_u \omega) \wedge \omega$. Therefore, $-(\iota_u \omega) \wedge \omega - u \wedge d(\iota_u \omega) = -(\iota_u \omega) \wedge \omega - d(u \wedge \iota_u \omega) + (\iota_u \omega) \wedge \omega = -d(u \wedge \iota_u \omega)$.

Combining, we obtain

$$\partial_t H = -d\left[\left(\frac{1}{2}|u|^2 + h\right)\omega + u \wedge \iota_u \omega\right]. \tag{B5}$$

*c. Conservation Law* Define the flux 2-form

$$J_H = \left(\frac{1}{2}|u|^2 + h\right)\omega + u \wedge \iota_u\omega. \tag{B6}$$

Then the helicity density local conservation law (B5) is $\partial_t H + dJ_H = 0$.

Integrating over a spatial domain $\mathscr{D}$ and applying Stokes' theorem, $\frac{d}{dt}\int_{\mathscr{D}} H = -\int_{\mathscr{D}} dJ_H = -\int_{\partial\mathscr{D}} J_H$. Under periodic boundary conditions or if fields decay sufficiently fast at infinity, the boundary integral vanishes, yielding conservation of total helicity: $\frac{d}{dt}\int_{\mathscr{D}} H = 0$. □

### 2. Comparison and "extension" of the helicity conservation law

We now offer the comparison of the proofs and a partial verification of the results from Chefranov & Chefranov's[12] presentation of the solution with Lagrangian mapping for the Burgers equation in the infinite domain.

*a. Khesin & Chekanov's proof* In their geometric approach, Khesin and Chekanov treated the incompressible ideal fluid as a Hamiltonian system on the group of volume-preserving diffeomorphisms. For incompressible flows, helicity emerges as a Casimir invariant of the Poisson structure, independent of the continuity equation (which is encoded in the volume-preserving constraint).

For barotropic flows, they geometrized the mass conservation as the transport of the density form, reducing the system to an incompressible flow with respect to the evolving volume form. Thus, their proof for barotropic flows indirectly incorporates mass conservation through this geometric reduction. This by itself is a beautiful and powerful treatment, especially for other more general purposes, but not necessary for the helicity conservation.

*b. Present proof* Our exterior calculus proof demonstrates that helicity conservation follows solely from the gradient form of the force term in the momentum equation, without invoking either the continuity equation or an incompressibility constraint. This makes the invariance manifestly independent of mass conservation and applicable/extendable to a broader class of flows, including the Burgers equation decoupled from the mass equation ($h = 0$) and our CWDRFs with modified continuity equations (the same calculations without any particular explicit specifications).

*c. Verification for the ideal solution to the Burgers equation in the infinite domain* Chefranov & Chefranov[12] provide the velocity and vorticity:

$$u_i[\boldsymbol{x}(\boldsymbol{\alpha},t),t] = u_{0i}(\boldsymbol{\alpha}),\ \omega_i[\boldsymbol{x}(\boldsymbol{\alpha},t),t] = \frac{\omega_{0i}(\boldsymbol{\alpha}) + t\,\omega_{0j}(\boldsymbol{\alpha})\frac{\partial u_{0i}(\boldsymbol{\alpha})}{\partial \alpha_j}}{\det A(\boldsymbol{\alpha},t)}, \tag{B7}$$

where $\boldsymbol{\alpha}$ denotes the Lagrangian position at $t = 0$, referred to by the index 0, and $A = \frac{\partial \boldsymbol{x}}{\partial \boldsymbol{\alpha}}$ in the ideal Burgers Lagrangian mapping $\boldsymbol{x} = \boldsymbol{\alpha} + t\boldsymbol{u}_0(\boldsymbol{\alpha})$. And we have the total helicity:

$$\begin{aligned} H(t) &= \int u_i\omega_i\, d^3x = \int u_{0i}(\boldsymbol{\alpha}) \cdot \frac{\omega_{0i}(\boldsymbol{\alpha}) + t\,\omega_{0j}(\boldsymbol{\alpha})\frac{\partial u_{0i}(\boldsymbol{\alpha})}{\partial \alpha_j}}{\det A} \cdot \det A\, d^3\alpha \\ &= \int u_{0i}(\boldsymbol{\alpha})\omega_{0i}(\boldsymbol{\alpha})\, d^3\alpha + t \int u_{0i}(\boldsymbol{\alpha})\omega_{0j}(\boldsymbol{\alpha})\frac{\partial u_{0i}(\boldsymbol{\alpha})}{\partial \alpha_j}\, d^3\alpha \\ &= H(0) + t \cdot B, \end{aligned} \tag{B8}$$

where

$$B = \int u_{0i}\omega_{0j}\frac{\partial u_{0i}}{\partial \alpha_j}\, d^3\alpha = \int \frac{\partial}{\partial \alpha_j}\left(\omega_{0j}\frac{1}{2}u_0^2\right) d^3\alpha - \int \frac{1}{2}u_0^2\frac{\partial \omega_{0j}}{\partial \alpha_j} d^3\alpha. \tag{B9}$$

The first term in the above right-hand side, from integration by parts, transforms into boundary integration by the Stokes theorem and vanishes with the fields decaying fast enough at infinity, and the second term is 0 due to the incompressibility of vorticity, thus the helicity invariance $H(t) = H(0)$.

Actually, the second line of the above Eq. (B8) can be obtained directly from the intermediate Eq. (5.3) of Ref. 12 without the necessity of introducing the Jacobian $\det A$. But, the above derivation appears more convenient for us, since Eq. (B7) is intuitively easy to understand, with the inviscid Burgers dynamcis being the vanishing of the material derivative of the velocity, $D_t\boldsymbol{u} = 0$, and with their vorticity expression being seen to be the apparent generalization of the familiar Cauchy formula[16] for an ideal incompressible Euler flow; otherwise, it would take more space for us to explain their Eq. (5.3) to which we refer the interested audience to Ref. 12.

## REFERENCES

[1] K. Julien and E. Knobloch, "Reduced models for fluid flows with strong constraints," J. Math. Phys. 48, 165405 (2007).

[2] J.-Z. Zhu, "Vorticity and helicity decompositions and dynamics with real Schur form of the velocity gradient," Phys. Fluids **30**, 031703 (2018).

[3] J.-Z. Zhu, "Thermodynamic and vortic structures of real Schur flows," J. Math. Phys. **62**, 083101 (2021).

[4] J.-Z. Zhu, "Real Schur flow computations, helicity fastening effects and Bagua-pattern cyclones," Phys. Fluids **33**, 107112 (2021).

[5] E.g., R. A. Horn & C. R. Johnson, Matrix analysis, Cambridge University Press (1990).

[6] C. J. Keylock, "Studying turbulence structure near the wall in hydrodynamic flows: An approach based on the Schur decomposition of the velocity gradient tensor," Journal of Hydrodynamics **34**, 806 (2022).

[7] Z. Li, "Schur Forms and Normal-Nilpotent Decompositions," Applied Mathematics and Mechanics, **45**(9), 1200 – 1211 (2024).

[8] E.g., V. I. Arnold, Ordinary Differential Equations (3rd ed.) Springer (1992).

[9] J.-J. Moreau, "Constantes d'un îlot tourbillonnaire en fluide parfait barotrope," *Comptes rendus hebdomadaires des séances de l'Académie des sciences* **252**, 2810–2812 (1961).

[10] B. Khesin & Yu V. Chekanov, "Invariants of the Euler equations for ideal or barotropic hydrodynamics and superconductivity in $D$ dimensions," Phys. D **40**, 119–131 (1989).

[11] J.-Z. Zhu, "Structures and Dynamics of Lone Schur Flows with Vorticity but no Swirls", preprint (2021: https://doi.org/10.48550/arXiv.2110.02225); J.-Z. Zhu, "On dimensionally reduced flows and anisotropic helicity polarization effects", preprint (2021) .

[12] S. G. Chefranov & A. S. Chefranov, "Exact solution of the compressible Euler–Helmholtz equation and the millennium prize problem generalization," Phys. Scr. **94**, 054001 (2019).

[13] S. G. Chefranov, "An exact statistical closed description of vortex turbulence and of the diffusion of an impurity in a compressible medium," Sov. Phys. Dokl. **36**, 286–9 (1991).

[14] V. I. Arnold & B. A. Khesin, Topological methods in hydrodynamics. Springer (1998).

[15] S. Weinberg, Gravitation and Cosmology. John Wiley & Sons (1972).

[16] See, e.g., G. Eyink, "Stochastic line motion and stochastic flux conservation for nonideal hydromagnetic models," J. Math. Phys. **50**, 083102 (2009), and, N. Besse and U. Frisch, "Geometric formulation of the Cauchy invariants for incompressible Euler flow in flat and curved spaces," J. Fluid Mech. **825**, 412 – 478 (2017), for different generalizations.